\documentclass[runningheads]{llncs}
\usepackage{pgfplots}
\pgfplotsset{compat=1.18} 

\usepackage{enumitem}

\usepackage{adjustbox}
\usepackage{float}
\usepackage{multirow}
\usepackage[section]{placeins}
\usepackage{pgf-pie}  
\usepackage{array}
\usepackage{xurl}
\usepackage{amssymb}

\definecolor{ibmyellow}{HTML}{ffb000}
\definecolor{ibmorange}{HTML}{fe6100}
\definecolor{ibmmagenta}{HTML}{dc267f}
\definecolor{ibmindigo}{HTML}{785ef0}
\definecolor{ibmblue}{HTML}{648fff}

\usepackage{hyperref}
\hypersetup{
    colorlinks=true,
    linkcolor=ibmmagenta,
    filecolor=ibmorange,      
    urlcolor=ibmblue,
    citecolor=ibmmagenta,
    }

\usepackage[T1]{fontenc}
\usepackage{graphicx}
\usepackage{color}

\begin{document}
\title{Privacy is Fungibility: Why Endogenous Tokens Are Not Money}
%
%
\author{
Alex Lynham\inst{1}
\and
Geoffrey Goodell\inst{1}
}
\authorrunning{A. Lynham and G. Goodell}
%
\institute{UCL, 66-72 Gower St, London WC1E 6BT}
\maketitle              
\begin{abstract}

In this paper, we make a case that endogenous tokens such as cryptoassets are not money. First, we define and classify tokens found on public, permissionless ledgers, contrasting them with privately issued stablecoins and proposed CBDC designs. We then discuss the work of Kahn et al in \textit{Money is Privacy} on cash versus simplified credit, and we extend their analysis to the situation found on most public, permissionless ledgers. Many public, permissionless ledgers utilize an account-based abstraction for balances, resulting in a default state that maps onto the most harmful models of agent interaction enumerated in \textit{Money is Privacy}. The conclusion is threefold: that most blockchain economies lack a cash-like primitive; that stablecoins do not intrinsically fulfil this role; and that the reliance of a network on an endogenous token for security exposes holders even of a privacy-preserving asset to the same risk, if that asset relies on the same global ledger state as the endogenous token.

\keywords{blockchain \and privacy \and credit \and tokens \and security}
\end{abstract}

\newcommand{\blockquote}{\medskip \noindent \leftskip 16pt}

\newcommand{\blockquoteend}{\medskip \leftskip 0pt}

\newcommand{\blockquoteendnoskip}{\leftskip 0pt}

\section{Introduction}

In this paper, we outline the requirements for a blockchain-based asset to be considered `money.' To do this, we build on the definition found in the \textit{Money is Privacy} paper by Kahn et al. on cash-like money and credit. Kahn et al. define money in terms of its privacy affordances: ``a money purchase does not identify the purchaser, whereas a credit purchase does\ldots [t]his property of money\ldots is of potential social value in economic situations where the parties in the transaction cannot trust each other not to take subsequent opportunistic actions.''~\cite{kahnetal}


We first classify the endogenous tokens found on public, permissionless ledgers, before then comparing the assets described in \textit{Money is Privacy} to the assets we have classified. 

The key contributions of this paper are:

\begin{enumerate}
    \item An analysis of ledger-based tokens in terms of their trust locus and security model. 
    \item An argument about the nature of the ``endogenous tokens'' found on most public, permissionless blockchains; simply put, they are \textit{not money},\footnote{Arguably, they are not even tokens, since most are not bearer instruments and exist only within an account abstraction held by an intermediary---for example, a ledger.} they are credit.
\end{enumerate}

Our argument for the second point is essentially as follows, building on Kahn et al.: if an intermediary can selectively revoke the asset, it is not money. If the asset is publicly observable, it is not money. If the asset is account-based,\footnote{That is, if its representation is as a key-value pairing denoting an address, or identity, even if arbitrary, and a balance. The key here is that an account always implies an intermediary; an \textit{account provider} provides an account to the \textit{account holder}.} as opposed to a fungible bearer token, it is not money.

In the Kahn paper, the continued fungibility of money, even in the case that it is stolen, is a core part of its definition. This requires privacy, if not outright obliviousness at a systemic level.\footnote{We argue that not just `privacy' as in Kahn, but `obliviousness,' that is, the property that all parties are required to \textit{not} know all states, is desirable.} As a necessary implication of this, there can be no intermediary---whereas in the case of blockchains, a ledger with ``infallibility of memory,''~\cite{kahnetal} the blockchain itself, its technical topology and its governance structure represent an intermediary, even if the chain data are private.

In Section 2 we examine `endogenous tokens,' that is, tokens whose security is endogenous to their site of representation,\footnote{We refer to this axis as `security locus' in our classification.} and whose regulation, or governance\footnote{We refer to this axis as `trust locus' in our classification.} is endogenous to the site of its current representation.\footnote{In our classification matrix, `endogenous tokens' are those in the top-left quadrant.} We then detail our classification of trust and security locus, and summarize some of the key arguments found in \textit{Money is Privacy}. 

In Section 3, we extend the analysis of Kahn et al. with limited cases covering the potential utility of theft. In Section 4 we analyse why this simplistic, but potentially damning, analysis likely does not imply a complete reversion to autarky or that collapse is inevitable, drawing on the work on economic security by Budish et al.,~\cite{budish2024economiclimitspermissionlessconsensus} as well as our prior papers on defining decentralization and conditional immutability on public, permissionless ledgers via qualitative fieldwork.~\cite{lynham2025decentralizationqualitativesurveynode,lynham2025definingdltimmutabilityqualitative} Finally in Section 5 we conclude.

Having classified tokens in terms of their trust locus and security model, we analyse them in light of the models of Kahn et al., finding that tokens with an endogenous trust model, but especially tokens with both an endogenous trust and security model map on to the simplified credit described by Kahn et al. This intuitively makes sense as they are revocable; their security model relies on ``collapse'' of the value of the asset's value to ensure its integrity, something described by Budish et al.,~\cite{budish2024economiclimitspermissionlessconsensus} and mirrored by a ``reversion to autarky'' in the credit models of Kahn et al.~\cite{kahnetal} Moreover, we note that an account-based model is more like a credit ledger than a mechanism for transactions with atomic tokens.

Finally, we further extend the analysis of Kahn et al to make some hypotheses about its application to blockchains. We conclude that if endogenous tokens map onto the simple credit mechanisms described in their models, then this implies that there is a missing cash-like primitive in these systems. We argue that most extant stablecoins do not fulfil this role, due to both their status in our classification and their lack of privacy affordances.\footnote{As well as their taking on the risk of any endogenous tokens that exist on ledgers that they rely on, but that expansion of scope is confined to Future Work.}

\section{Context}

\subsection{Endogenous Tokens}
{\renewcommand{\arraystretch}{1.5}
\quad{\setlength{\tabcolsep}{1em}
\begin{table}[t]
    \centering
    \caption{Exogenous and endogenous token security and trust locus}
    \label{table:token_classification}
    \begin{adjustbox}{width={\textwidth},totalheight={\textheight},keepaspectratio}
    \begin{tabular}{|p{0.10\linewidth}|p{0.14\linewidth}|p{0.38\linewidth}|p{0.38\linewidth}|}   
         \cline{3-4}
         \multicolumn{2}{c|}{}  & \multicolumn{2}{c|}{\textbf{Ledger Trust locus}}\\ 
         \cline{3-4} 
         \multicolumn{2}{c|}{} & Endogenous & Exogenous\\ 
         \hline 
         \multirow{4}{*}{\parbox{0.10\textwidth}{\centering\textbf{Token security}}} &  Endogenous &  Example: Typical Proof-of-Stake chain (e.g. Cosmos chain) with “object based”~\cite{Marple_2021} value backing \newline\newline\newline
Effect of attack: Rewrite causes collapse
& Example: Privately operated stablecoin with “claim-based”~\cite{Marple_2021} value backing, operated on private or consortium ledger. \newline\newline
Effect of attack: Operator or consortium retain control of finality subject to enforcement (e.g. regulation)
\\ \cline{2-4}
         &  Exogenous&  Example: Typical Proof-of-Work chain (e.g. Bitcoin) \newline\newline
Effect of attack: Rewrite causes collapse
& Example: Cash-like CBDC \newline\newline\newline
Effect of attack: Individual notes are hard to target, meaning they are fungible even if stolen (in the case of digital cash) and there is a system-level cost in switching off the system. Thus any attack is likely causally reversed, i.e. an attack on the currency exogenously results in effects on the CBDC.
\\ \hline 
    \end{tabular}
    \end{adjustbox}
\end{table}
}}%

\noindent To begin our analysis of trust and economic security, we consider the question posed by Poelstra, ``is it possible to obtain a distributed consensus without provably consuming some resource outside of the system? Intuitively, the answer is no.''~\cite{Poelstra2015DistributedCF}

To do this, we classify tokens in terms of their token security and trust locus. Token security relates to the token's value basis,\footnote{This is important since most public, permissionless blockchains rely on `economic security,' that is, incentives linked to token value, for their security model.~\cite{bitcoin_nakamoto}} while its trust locus concerns its integrity---or finality, in consensus system terminology.\footnote{As well as required technical affordances, it also concerns who is responsible for that integrity, meaning it interacts with protocol design, as well as governance and regulation.}

In this paper we primarily analyse endogenous tokens, which we define as tokens that are both endogenous in their token security and trust locus,\footnote{Note that strictly, trust locus is about the governance of the ledger on which the token currently exists.} the top-left quadrant in Table \ref{table:token_classification}. In this classification, token security is considered in terms of consumption of an outside resource---in other words, a concrete connection to entropy,\footnote{Consumption of CPU cycles, in the case of Proof-of-Work, or labour or economic product that generates taxable revenue, in the case of fiat currency.} while trust locus is whether there is anything outside of the ledger that guarantees its integrity---for example, a government, regulator, or other institution.\footnote{Of course, nothing is ever completely certain, so perhaps we might build on the discussion of `confidence' in our prior work,~\cite{lynham2025definingdltimmutabilityqualitative} after De Filippi et al.~\cite{DEFILIPPI2020101284} and state that rational, calculative agents should have confidence bordering on certainty over the integrity of the asset, now, and into the future.} Proof-of-Work and (Delegated)Proof-of-Stake tokens created on public permissionless blockchains would be expected to fall into the ``endogenous'' trust (left) column. 

Perhaps a useful heuristic is: when is the value of the token established? In the case of Proof-of-Work, the minting of tokens requires that the connection to entropy be ex-post (the work has already been done). However, any mechanism that operates ex-ante with regard to entropy, or has no connection to entropy at all, such as minting in most Proof-of-Stake systems, is different. It is not only endogenous in the sense of being intrinsic to the ledger on which it is cardinally represented, but also issued ex-ante (before its value is established, unlike a ``claim-based'' asset).\footnote{This is a crucial consideration, because absent value creation in the system, value needs be established elsewhere. A system where all assets are endogenous to the system creates a zero-sum game.}

In the case of money issued by a central bank, the bank is guaranteeing the face value of the money before issuance,~\footnote{The money being backed by the tax receipts of a state, as well as presumably an army, except in the case of Costa Rica.} while in the case of a ``claim-based''~\cite{Marple_2021} private stablecoin, its value is established at the point of redemption---it is a representation, not the thing itself, and relies on a relationship to an underlying currency or an asset, just as a pegged currency does. However, it is exposed to different risks and vectors for attack than the underlying currency or asset.

Since there is a direct connection between the representation of a token in its system of record, `Token security' in Table \ref{table:token_classification} might be equally thought of as, `Security model of the ledger on which the token \textit{currently} depends for its integrity.' The same is true of its trust locus. The token might be minted, or issued elsewhere, but its location at a point in time will change the locus of its trust and security, for example, if it is bridged to a different ledger.\footnote{A highly relevant example would be a stablecoin such as USDC, bridged to a different ledger.} 

It is important to note that in this classification, the same token in two different contexts might be in different quadrants. For example, a stablecoin within the rails of a private provider's network benefits from claim-based value backing to counterweight the fact that the finality of that ledger is likely not backed by the strict consumption of an outside resource (unlike, say, a Proof-of-Work ledger), and thus falls in the top-right quadrant. The same stablecoin bridged to a public, permissionless network, however, would fall in the top-left quadrant, as the private provider no longer retains control over finality with relation to that token. 

In Section 4 we will analyse this classification in light of ``expensive due to collapse'' property described by Budish, Lewis-Pye and Roughgarden~\cite{budish2024economiclimitspermissionlessconsensus} in more detail; for now it is enough to observe that although we primarily concern ourselves with Proof-of-Stake ledgers in the top left quadrant, that Proof-of-Work ledgers, while arguably having a different security model, in fact are vulnerable to the same failure mode, where rewrite causes collapse.\footnote{If trust is undermined. For further analysis of this trust-contingent immutability, see our prior paper.~\cite{lynham2025definingdltimmutabilityqualitative}} In our analysis, this collapse is a feature of endogenous trust in the ledger,\footnote{i.e. endogenous governance.} and simply consuming “some resource outside of the system”, in this case large amounts of CPU time, does not result in greater security.

\subsection{Money is Privacy}

In their paper \textit{Money is Privacy},~\cite{kahnetal} Kahn et al. model several different types of agentic actions, transacting with both a credit-like instrument and a money-like instrument. Most relevant for our analysis is their model 3.3, \textit{Money as an Alternative to Full Information}, but they document several interesting trade-offs, and we summarize them in Table \ref{table:money_is_privacy}.

Their definition of money requires that it is anonymous, as a check against opportunistic actions by counterparties. They note, ``that in addition to its value as a possibly imperfect proxy for credit, money also derives value from its use in anonymous exchanges, facilitating certain otherwise infeasible transactions.''~\cite{kahnetal} In other words, anonymity has a benefit in terms of improving transaction-cost economics.

Where ex-post attacks on integrity might be expected, anonymity has a value. ``The point is in fact of wider applicability. Monetary economies are examples of infinitely repeated games of imperfect monitoring, because individuals' monetary holdings and monetary transactions are private information. In general the imperfection of the monitoring restricts the equilibrium set of outcomes in such games\ldots However, when privacy of transactions reduces opportunism, it expands the equilibrium set.''~\cite{kahnetal}

Finally, they note that their model requires ``infallibility of memory,'' and that it is this property ``that renders [memory] exploitable by outside parties.'' The parallel here to a public, permissionless blockchain with visible history should not need further explanation. Theft of the transacted good is the only moral hazard they analyse, but others might be expected, including other forms of theft and exploitation arising from disclosure of transaction data;\footnote{We discuss the case of MEV, or Maximal Extractable Value, which often results in user harm, in Section 4.1.1.} ``What is key to our analysis is the possibility of ex post opportunistic behavior that may arise under limited enforcement.''~\footnote{From our prior work,~\cite{lynham2025definingdltimmutabilityqualitative} we might argue that many blockchains evidence emergent regulation; nevertheless it would be charitable to describe this as anything other than `limited.'}~\cite{kahnetal}

{\renewcommand{\arraystretch}{1.5}
\quad{\setlength{\tabcolsep}{1em}
\begin{table}[t]
    \centering
    \caption{Models in \textit{Money is Privacy}}
    \label{table:money_is_privacy}
    \begin{adjustbox}{width={\textwidth},totalheight={\textheight},keepaspectratio}
    \begin{tabular}{|p{0.15\linewidth}|p{0.85\linewidth}|}   
         \hline
         \textbf{Model} & \textbf{Summary} \\
         \hline 
         \raggedright Model 2.2, Bilateral Information on Goods Received & ``In this environment, supplier in an exchange is informed of the identity of the recipient. Each potential supplier must decide whether the disutility of production is worth the gain of `reciprocal privileges' with a given consumer.''~\cite{kahnetal} \\
         \hline 
         \raggedright Model 2.3, Full Information on Goods Received & ``there is full information about the identity of recipients of goods in any exchange that occurs. A failure to supply is a publicly observable event; thus we examine equilibria in which a failure to supply leads to autarky.''~\cite{kahnetal} \\
         \hline 
         \raggedright Model 3.1, Money under Semianonymity & ``Suppose that would-be consumers cannot (or prefer not to) reveal their identity to would-be suppliers. Trade in this case is `semianonymous,'' since the supplier's identity is always known to the consumer. Money offers opportunities for exchange in this environment.''~\cite{kahnetal} \\
         \hline 
         \raggedright Model 3.2, Money as an Alternative to Bilateral Information & ``we allow [agents] the following two choices: they may anonymously purchase goods with money or they may choose to reveal their identity to their suppliers (and no one else) with the intent of obtaining `credit' for future reciprocal actions\ldots As was the case in the previous section, a credit purchase exposes the purchaser to the possibility of theft from the supplier.''~\cite{kahnetal} \\
         \hline 
         \raggedright Model 3.3, Money as an Alternative to Full Information & ``consumers have a choice between using money or revealing their identity in credit transactions. In the latter case their identity is revealed not only to their counterparty, but also to all other agents in the economy. We assume if a supplier refuses to supply in a credit transaction, his identity is revealed, and the supplier is reduced to monetary transactions only. In monetary transactions, identities are not revealed.''~\cite{kahnetal}\\
         \hline 
    \end{tabular}
    \end{adjustbox}
\end{table}
}}
Thus, their bar for a consumer being at risk of theft is low, just that they identify themselves. In model 2.2, \textit{Bilateral Information on Goods Received} the risk comes from the vendor. In 3.3, \textit{Money as an Alternative to Full Information} it comes from all other agents, akin to Liu Cixin's `Dark Forest,' where to announce one's presence results in death.~\cite{darkforest}

Kahn et al. argue that cash is useful in low-trust situations. This describes the de facto situation on public, permissionless ledgers.\footnote{If opportunism were easily accessible to agents, a point we will discuss further in Section 4. Briefly, a protocol-level constraint on double spending combined with non-protocol factors like validator set integrity, a high enough market cap to increase the cost of an attack beyond the means of a single entity, and multiple development teams mitigate the risk of opportunistic actions to manipulate the ledger and revert transactions.~\cite{buterinsettlementfinality,ethereumvalidatorchecklist}} They observe, ``if the cost and likelihood of theft are too high, then only exchange with money is possible.'' 

This is problematic when we consider that cryptoassets map best not to \textit{money} in their model, but to their simplified version of \textit{credit}. Following the argument of Budish et al., the ``cost'' is potentially total collapse, so the question becomes, what kind of theft triggers this sanction? 

Kahn et al. define this simplified credit in the introduction to model 3.2, \textit{Money as an Alternative to Bilateral Information},

\blockquote
``[Agents] may anonymously purchase goods with money or they may choose to reveal their identity to their suppliers (and no one else) with the intent of obtaining `credit' for future reciprocal actions\ldots As was the case in the previous section, a credit purchase exposes the purchaser to the possibility of theft from the supplier.''~\cite{kahnetal}

\blockquoteend

Defections, that is, failures to supply, result in trust collapse, and a reversion to cash-only transactions. This they compare to a definition of money, which acts as a way of preventing loss by adding consumer privacy,

\blockquote
``a unit of money [is] an indivisible, inherently valueless, noncounterfeitable object, where each agent can hold a maximum of one such object. There is a fixed supply of money and not all agents possess money at any given time. Money confers anonymity: A consumer making a purchase with money does not reveal his identity to his supplier or to others.''~\cite{kahnetal}

\blockquoteend

Although discourse around cryptocurrencies tends to frame them as something closest to cash, the reality is that their closest analogue in Kahn et al's model is this definition of credit. This holds true for anything that is account-based, where balances can be publicly seen, but it also affects non-private UTXO-based coins where there is an implicit address and balance pair.

As venture capital pours into the blockchain indexing space, with leading firms like Chainalysis raising \$170M against an \$8.6bn valuation,~\cite{reuterschainalysis} this situation will become the rule in any public ledger, regardless of specifics of state management. In addition, over time, pattern analysis will likely make the adoption of multiple accounts or addresses less effective against surveillance of holdings.

\section{Theft}

We begin by considering a simplified model for theft in a situation where all agents can view all other agents' balances. This is a reframing of Kahn, akin to the situation found on a public, permissionless ledger with account-based abstractions for balance state, such as that found on Ethereum, or in the Cosmos Ecosystem, the networks primarily under study in our past papers.~\cite{lynham2025decentralizationqualitativesurveynode,lynham2025definingdltimmutabilityqualitative}

Although models involving money (3.1-3.3) more explicitly model the interaction between credit and money, cryptoassets have a token ``form'', and as such it is useful to try and apply the theft-of-goods model in 2.3, \textit{Full Information on Goods Received} to cryptoassets. What this will show is that infallibility of memory, observable holdings and observable identity are as problematic as the reader would intuit.

Kahn et al advance a simplified model for theft, stating, ``[N]o robber will attempt to steal from an individual at random.'':~\cite{kahnetal}

\begin{equation}
\label{theft_model}
\frac{\alpha\varepsilon f}{N-1} < c
\end{equation}

Where:

\begin{enumerate}
    \item \(u\) is utility of the good to the consumer
    \item After a theft of \(f\) utils, \(u - f\) is the cost to the victim of the theft
    \item \(\varepsilon f\) is the benefit to the robber from a successful theft (where \(0 < \varepsilon < 1\)) 
    \item A theft attempt costs the robber \(c\) utils
    \item \(N\) is number of agents
    \item A theft attempt will be successful a fraction \(\alpha\) of the time iff the victim possesses the good
    \item \(\frac{\alpha}{(N - 1)}\) is the chance of successful theft by choosing another agent’s location at random
    \item All agents are risk neutral and have a common discount factor, \(\delta\)
\end{enumerate}

It is important to note that while we build on this model to show risk of theft of cryptoassets, in practice, we hypothesise that the difficulty of opportunistic attack on ledgers makes thefts against individual agents unlikely. Additionally due to the ontological properties of any cryptoassets on ledgers that implement an account-to-balance mapping,\footnote{We in general refer to this design pattern as an Account model in this paper.} we caveat in advance that discussion in the context of model 3.3, \textit{Money as an Alternative to Full Information} is also required. We do not attempt to extend model 3.3, \textit{Money as an Alternative to Full Information} because its utility assumptions are---in the simplified case, we believe---compatible with our argument (Section 4). That is not the case with 2.3, \textit{Full Information on Goods Received} mapped to cryptoassets, so we make an attempt to reconcile the model below. 

However, the model proposed by Kahn et al. only allows for a single unit of goods to be held. Since ``collapse'' is the only mechanism to penalize attacks,\footnote{In previous papers, we have focussed on ledger rewrites; here our attack is a theft from an agent, either executed on-chain (an unlikely scenario), or more likely via social engineering or even physical attack, as described in Section 4.} we hypothesise that this is unlikely to be triggered when a single agent is attacked, since trust in the governance of the ledger and legitimacy of ledger state has not been undermined.~\cite{lynham2025definingdltimmutabilityqualitative} This will be analysed further in Section 4.


 

\subsection{Extending the model for \textit{Full Information on Goods Received}}

To model a credit collapse---that is, the collapse in confidence of an endogenous token, we build on their model for \textit{Full Information on Goods Received} and use their model for goods as our stand-in for cryptoassets. In this model we envisage agents exchanging cryptoassets, with some attempting to steal assets.

A failure to supply results in a reversion to autarky, while some agents will attempt to rob goods. In their model, not all agents seek to exchange goods during a given cycle, or day, and only one robber exists per cycle. The simplification of the robber mechanic works for our analysis, while since we conceptualize the supplier-consumer relationship as sending cryptoassets instead of demanding and supplying a good, it is similarly reasonable to not expect all agents to transact every cycle.

As we confine ourselves to goods like their model, we note that while \(u\) and \(f\) are in most cases equal (see the breakdown of \ref{theft_model} in Section 2.2), there is some subtlety. Thus Upsilon, \(\upsilon\), replaces the robber's expected benefit (\(\epsilon f\)) for a theft.

We assume that instead of a model like Kahn et al. where the potential utility after a theft is as in (\ref{theft_instantaneous_benefit}) in terms of goods, in our approximation it is either \(u\)\footnote{i.e. 1 if asset holdings are limited to 1.} or 0. If the ledger does not collapse, it is \(u\); if the ledger collapses, it is 0. Thus:

\begin{equation}
\label{upsilon_definition}
\upsilon = u\veebar0
\end{equation}

Equivalently:

\begin{equation}
\label{upsilon_definition_sets}
\upsilon \in \{u,0\}
\end{equation}

Moreover, since all network participants are harmed in the case of collapse, an attacker might not be deterred by the cost being less than or equal to zero (\ref{agents_supply_sustainable_equilibrium_1}). In the event that an attacker can be stopped by a targeted rewrite, freeze, or seizure, their likelihood of success ($\alpha$) would diminish, but as mentioned, this might not deter an attempt.

We begin by modelling the theft of a single unit of cryptoassets, assuming the target has the asset:

\begin{equation}
\label{theft_single_unit}
\frac{\alpha \upsilon}{N-1} > c
\end{equation}

Then the instantaneous expected benefit to theft of a good:
 
\begin{equation}
\label{theft_instantaneous_benefit}
\alpha \upsilon - c
\end{equation}

This means that theft will occur so long as the expected value of a stolen asset, multiplied by its likelihood of successful attack, exceeds the cost of carrying it out, since there is no longer reason to consider the likelihood that a victim has the asset. This is problematic because token holdings on a public ledger much more closely map to Kahn et al.'s model for credit than for cash, meaning that identity is revealed when transacting. In their model 2.3, \textit{Full Information on Goods Received}, this results in the situation that full information (that is, a situation where all identities are known, and theft exists) is only sustainable without theft when:

\begin{equation}
\label{theft_sustainable_equilibrium}
\alpha \varepsilon f - c \leq 0
\end{equation}

Meanwhile, full information is not sustainable without theft unless c is sufficiently negative:

\begin{equation}
\label{theft_sustainable_equilibrium}
\alpha u - c \leq 0
\end{equation}

We argue that in this case, the cost of theft, \(c\), is strictly equal to the attacker's ``stock value''~\cite{budish2024economiclimitspermissionlessconsensus} as described by Budish et al. We include \(s\), the cost of supply, which in most cases will just be any network transaction fees. 

Putting this together, under full information, a pure credit equilibrium, with theft, in which all agents transact using tokens is sustainable if and only if:

\begin{equation}
\label{agents_supply_sustainable_equilibrium_1}
\alpha \upsilon - c \geq 0
\end{equation}

\begin{equation}
\label{agents_supply_sustainable_equilibrium_2}
u - \alpha f - s + \alpha \upsilon - c \geq 0
\end{equation}

and

\begin{equation}
\label{agents_supply_sustainable_equilibrium_3}
\frac{u - \alpha f - s + \alpha \upsilon - c}{\delta N} \geq s - \frac{\alpha \upsilon}{N-1}
\end{equation}

We have here included \(f\), because the cost to the supplier is not formally the same as the benefit to the robber (even if the two are likely to both be \(u\) in many cases). Ignoring the complexity of such theft on an opportunistic basis (which probably renders \(\alpha\) vanishingly small),\footnote{This is due to double-spend prevention on the one hand, and the fact that as part of a consensus mechanism, all validators or miners need to agree on a block.} the expected value of exchange is:

\begin{equation}
\label{value_of_exchange}
\frac{u - s - \alpha \upsilon - c}{\delta N}
\end{equation}

As in their model, there is a condition that theft is tempting (\ref{agents_supply_sustainable_equilibrium_1}); the next requirement for exchange to take place is that ``the expected value of [exchange] is positive.'' It is this condition that means agents choose to transact. With the introduction of $\upsilon$ (\ref{upsilon_definition}) we can see that the expected value of a transaction is all-or-nothing---either the recipient captures the full utility \(u\), or the robber receives the full utility---this is why protocol-enforced security against opportunistic actions is so important.

\subsection{Extending the model for \textit{Money as an Alternative to Full Information}}

Briefly touching on the models including money, we might suggest that the instantaneous expected benefit of the theft of a single unit of money would be its full expected utility, \(u\), rather than \(\varepsilon f\), were it to be fungible enough to exchanged for a good, i.e.

\begin{equation}
\label{theft_single_unit_3_3}
\alpha u - c
\end{equation}

Unlike in the formalizations from the previous section (\ref{theft_single_unit}), the non-observability of a unit of money means it retains its fungibility even if stolen, meaning the constraints in the prior section (\ref{upsilon_definition}) do not apply.\footnote{While cash can, for example, be removed from circulation by a central bank, all cash needs to be voided or blocked and a single holder cannot be targeted. That is, controls are system-wide, and costs are system-wide.} 
Following this, and especially were sellers able to hold more money, the incentive toward theft is increased. Formally, because of public visibility, the guesswork is taken out of identifying a target, and the likelihood of a successful attack, \(\alpha\), is the only factor preventing a theft.

Even so, the risk-reward is a probability game---a would-be thief is running the risk of the ledger rewriting the theft, however slight, if the theft is large enough.\footnote{As indeed occurred following the Ethereum DAO hack. In the case of a true cash-like asset, the only likely remediation would be reverting to a prior state and resuming---which would be very unlikely. This is another way that obliviousness (in the case of cash) makes rewrite events less likely.} 

However, what if, as in our blockchain example, all balances were known by all agents in the economy?\footnote{An example here could be a public, account-based stablecoin that was not oblivious.} In that case, the risk-reward would be clearly known, as in model 2.3, \textit{Full Information on Goods Received}. The result is that the statements about theft we made in the prior section (Section 3.1) would hold, and the asset could not be deemed cash-like, or indeed ``money'' in the sense used by Kahn et al. 

Recall that the instantaneous expected benefit to theft of a good is:

\begin{equation}
\label{theft_instantaneous_benefit}
\alpha \varepsilon f - c 
\end{equation}

Most problematically, \(c\) must be sufficiently large to deter an opportunistic theft.\footnote{For the moment, in this hypothetical scenario, we ignore the fact that opportunistic attacks requiring technical knowledge are difficult to execute. The reality will be analysed in Section 4.} In permissionless DLT systems however, \(c\) is likely to be very small. As we have discussed previously~\cite{lynham2025decentralizationqualitativesurveynode} and others have observed, unless an attacker loses their ``stock value,''~\cite{budish2024economiclimitspermissionlessconsensus} the cost of an attack, especially a social engineering attack, is likely to be ``cheap'' or even zero. Indeed, this quality of being a ``cheap'' attack scales to an attack on an entire network.~\cite{Houy2014ItWC}

Overall, what this shows is that it is the property of privacy, or, in our argument, the stronger term, `obliviousness' that distinguishes money from different types of asset. Remove obliviousness, and even a stablecoin with some ostensibly `money-like' properties, such as an exogenous trust locus (see Table \ref{table:token_classification}) has more in common with a cryptoasset with endogenous trust and integrity.

\subsection{The Cost of Attacks}

As observed before, \(c\) is strictly equal to the attacker's ``stock value''~\cite{budish2024economiclimitspermissionlessconsensus} in the Budish et al. conceptualization, but if the attack was designed to cause economic damage (e.g. executed by a competitor or an agent that stood to benefit economically from a token collapse, as in the case of the Terra UST stablecoin collapse~\cite{briola_terra}) then it might be that there is a utility gained by an attacker that is exogenous to the chain under study, such that both exogenous utility, \(u_e\) and exogenous cost \(c_e\) have to be considered.\footnote{In Budish~\cite{budish10.1093/qje/qjae033} this is discussed. The argument is made, ``Nakamoto trust is incentive compatible against an outsider attack, on a gross-cost basis, if the gross cost of attack exceeds the benefits of attack,'' but he goes on to demonstrate that the net cost and benefits are affected by other factors. His formalization is different but somewhat analogous to what we describe here.} 

This is not covered in the Kahn et al. model but could be formalized in isolation. We will briefly conduct the thought experiment, but leave any further developments for future work.

The instantaneous expected benefit to a theft or attack for a would-be robber is:

\begin{equation}
\label{theft_instantaneous_benefit_exogenous}
\alpha (\upsilon + u_e) - (c + c_e)
\end{equation}

Meanwhile, full information is not sustainable without theft unless \(c\) and \(c_e\) are sufficiently large:

\begin{equation}
\label{theft_sustainable_equilibrium}
\alpha (\upsilon + u_e) - (c + c_e) \leq 0
\end{equation}

As described previously, a final development of this scenario is that where holdings could be greater than 1, it may be that after a certain point, a holding \(n\) becomes big enough that a network would rewrite to revert a hack.\footnote{See Appendix B, Response 2 in our prior paper,~\cite{lynham2025decentralizationqualitativesurveynode} the historic post-DAO hack rewrite on Ethereum, or the tombstone reversions in the Cosmos Ecosystem we have documented previously.} Additionally we hypothesise some accounts are more likely to be corrected via rewrite, for example a chain's Community Pool, or the holding of a Foundation. We note that this is a function of an account-based model that requires interacting with revealed identity. 

However, for a single agent, with holdings greater than one, we need to adjust our formalization. The utility to an attacker of any theft, Mu (\(\mu\)), however executed, remains either \(0\) or \(tu\), where \(t\) is the total number of units stolen and \(u\) is the full utility of a single unit of goods, stolen by theft. Similarly to the definition of Upsilon in (\ref{upsilon_definition}) because the network collapses, or the network rewrites the theft, the full utility is retained, or nothing:

\begin{equation}
\label{theft_greater_than_1_instantaneous_benefit_exogenous}
\mu = tu \veebar 0
\end{equation}

We do not allow for the possibility of partial success. Note that, as defined, \(c\) is the cost to the robber of the attempted theft, and is not per-unit. Factoring in cost, the instantaneous expected benefit becomes:

\begin{equation}
\label{theft_greater_than_1_instantaneous_benefit_exogenous}
\alpha (\mu + u_e) - (c + c_e)
\end{equation}

There are in effect only three scenarios for the thief:

\begin{enumerate}
    \item The attack is rewritten or reverted, resulting in a \(\mu\) of 0, but potentially a positive \(u_e\).
    \item The ledger collapses, resulting in a \(\mu\) of 0, but potentially a positive \(u_e\).
    \item The ledger does not collapse, and the attack is not rewritten or reverted, resulting in the full utility of the attack (\((\mu + u_e) - (c + c_e)\)) being retained. In this case, there is the possibility that \(u_e\) might be zero, if the only goal of the attack was to collapse or disrupt the ledger.\footnote{For example in a griefing attack, such as the chain-halting cyberattack seen on the Juno blockchain during its Proposal 16 governance dispute.~\cite{coindeskjunohack}}
\end{enumerate}

These exogenous factors can have unpredictable results. For example, in the case of the Ethereum DAO hack, the endogenous benefit of the theft (\(\mu\)) was 0, as the Ethereum chain hard-forked to rewrite the hack. However, the creation of the Ethereum Classic blockchain with unaltered history resulted in an unexpected exogenous benefit. Specifically, the attacker received the value of the equivalent quantity of Ethereum Classic (ETC) tokens (3.6 million tokens). The hacker withdrew these on or around July 25, 2016. At that point ETC was trading for \$0.6024,~\cite{etc_price} meaning the value of \(u_e\) was around \$2,168,640, while \(c\) was likely nothing more than the cost of gas on Ethereum, and \(c_e\) the cost of gas on Ethereum Classic.


\section{Analysis}

\subsection{Why is theft not more common?}

If the previous theft models hold, then the obvious question is why theft is not more commonplace. Here the answer is simple. Unlike in the abstract model, where theft is technically simple and opportunism is trivially possible for agents, attacks on blockchains typically require highly specialised knowledge or a zero-day vulnerability to execute.\footnote{As we noted in saying \(\alpha\) was likely very small.} This reduces the opportunity for a would-be thief to engage in opportunistic theft.\footnote{Even so, loss events as a result of opportunistic action, for example via `sandwich attacks', where ``ill-intentioned actors exploit price slippage by positioning their transactions strategically around a target’s order to reap unfair profits,''~\cite{liangetal} do occur, and will be discussed in Section 4.1.1. One of the most dramatic of these incidents came in 2025 when a user swapping stablecoins was attacked, receiving only 5271 USDT in exchange for 220,764 USDC.~\cite{sandwich_attack} Per the Kahn definition, if these two assets were money, this attack could not have taken place.}

\subsubsection{4.1.1 MEV}

MEV can be used to provide examples of opportunistic actions by agents that results in loss of utility for the user and utility transfer to the attacking agent, but nevertheless fall short of outright theft. For this reason it is instructive to analyse in light of the Kahn models.

MEV, or Maximal Extractable Value, is the blanket term for various kinds of block transaction ordering.~\cite{fca_mev} Many of the most well-known examples, such as MEV-Geth for Ethereum\footnote{Now known as MEV-Boost.} work on a sealed-bid auction basis. Most benignly, this can be used to ensure a user's transaction makes it into a block. However, attacks using the technology are possible, such as frontrunning, backrunning, and sandwich attacks,~\cite{fca_mev} which can cost Decentralized Exchange (DEX) users large sums of money.~\cite{sandwich_attack}

Recalling the instantaneous benefit for a theft (\ref{theft_instantaneous_benefit}), we can see a parallel:

\begin{enumerate}
    \item $\alpha$, the likelihood of an attack succeeding, can be brought to 1 by a successful bid for block space.
    \item $\epsilon f$, the benefit to the attacker, can be calculated ahead of time due to public visibility of transactions in the mempool.
    \item $c$, the cost to the attacker, in the simple case is simply transaction costs (i.e. gas) plus the cost of the successful bid.
\end{enumerate}

Following this, the attacker's rationale for executing an attack is a simple calculation, nearly mirroring the conditions by which full information without theft is sustainable (\ref{theft_sustainable_equilibrium}).

\begin{equation}
\label{mev_sustainable_equilibrium}
\alpha \varepsilon f - c > 0
\end{equation}

The intuition here therefore is that this situation is not sustainable without theft, and indeed it is not. Our argument implies that such an avenue for opportunistic, zero-sum actions would prove deeply problematic, and so it has proved. MEV attacks are enough of an issue that many blockchains have introduced shielded mempools, such as Aptos, while blockchains devoted entirely to the shielded DEX concept, such as Penumbra, have also emerged.

For blockchains without a shielded mempool, the answer in many cases to why attacks do not happen on a constant basis---particularly in the DEX case, where users are most vulnerable to sandwich attacks---is simply that DEX user interfaces typically contain a slippage control. In addition to protecting users from the inherent volatility of an AMM, this control means their transaction will be cancelled if exploited, which means only large transactions are typically worth attacking to extract value.

\subsubsection{4.1.2 Outcome Theft}

Elsewhere, the risk of physical loss, extrapolated in the same way as Kahn et al.'s model, perhaps by the ``five dollar wrench attack''~\cite{xkcd} resulting in the loss of private keys, is small in the real world. Still, it is reported to be on the rise, with several high-profile incidents reported by news outlets.~\cite{parisattemptedkidnapping,pariskidnapping,sun2022,franceledgerattack,physicalattacksnyckidnapping} A community repository of physical attacks on Bitcoin holders has reported over two hundred in eleven years.~\cite{physicalbitcoinattacks}

Targeted attacks to steal private keys without physical coercion also happen, usually via social engineering, or SIM swapping,~\cite{simswap} and do succeed. Various figures are thrown around for the value of cryptocurrency hacks, with Chainalysis claiming their value as \$3.7 billion in 2022 and \$1.7 billion in 2023. Though Chainalysis note that the absolute value fell, they show the number of attacks increased. The real number is likely substantially higher, since their numbers are an approximation.~\cite{chainalysis2024}

Permissionless, public ledgers do provide security guarantees against opportunistic theft, an acceptable bar of immutability for many users that we termed \textit{Practical Immutability} in a prior paper.~\cite{lynham2025definingdltimmutabilityqualitative} This means that whatever the credit-like properties of these cryptoassets, the threat for most users from other agents' opportunism is low outside of the most low-effort cybercrime such as phishing.~\cite{phishingattacks} Still, these targeted attacks are only possible because users’ identities and holdings have been revealed by interacting with a public ledger.\footnote{Outside of the targeted attacks just discussed, `Spearfishing' large holders and institutions and targeting them for determined attacks is also commonplace.} Though Chainalysis and its competitors primarily market to governments and law enforcement, in future their data and tooling could conceivably become available to the general public. This would lead to an ironic situation where their data are most valuable either to law enforcement, or to criminals themselves. 

Unlike in Kahn et al.'s model, the granular withdrawal of credit facility on a per-user basis is less likely on a blockchain. This could be implemented via a blocklist or transaction filtering mechanism, or the tokens could be rendered non-fungible to the owner by brute seizure---for example, by moving the tokens from a wallet via rewrite to a smart contract wallet with limited functionality. This is unlikely to occur both for technical and ideological reasons, but chiefly it is unlikely because taking action might negatively impact trust in the governance of the ledger. As a result, targeted rewrites of balances, akin to a revocation of credit to a single agent in the Kahn models, are rare.\footnote{We have noted instances of rewrites in the Cosmos Ecosystem in prior work. For now, it is sufficient to observe that targeted seizures do occur, most famously in the case of Juno Proposal 16.~\cite{mintscanjunoprop16,coindeskprop16watershed}}

Due to the required technical sophistication of attacks against ledger state, targeted rewrites against ledger state, or withdrawal of credit for a single user or wallet, to put it another way, tend to occur via the governance topology of a network, occurring at the granularity of the network rather than of an entity attacking another entity. Budish et al., have described how economic security for blockchains with endogenous tokens is an ex-post mechanism that punishes all participants equally,~\cite{budish2024economiclimitspermissionlessconsensus} and that only this cost renders an attack expensive to an attacker. Thus, what we should expect is that most attacks executed exogenously without modifying ledger state, for example social attacks via network governance, do not trigger this mechanism, as they do not undermine trust in either the legitimacy of ledger state, or the network's governance topology.~\cite{lynham2025definingdltimmutabilityqualitative}

Indeed, we have described that many rewrites executed via the governance topology of a network do not trigger this security mechanism. We hypothesise, based on thematic analysis of fieldwork, that agents on a ledger operate in context of institutional trust where a mixture of calculative and non-calculative incentives combine to form the pragmatism that leads few to exit even in the case of rewrite. This means that it is hard to reason about the robustness of a ledger ex-ante, since when this endogenous institutional trust is destroyed, the Budish et al. ``collapse'' must necessarily follow.~\cite{lynham2025definingdltimmutabilityqualitative}

If this mechanism is triggered, then the result is similar to the ``reversion to autarky'' described by Kahn et al. Indeed, since most ledgers currently only have an endogenous token, such a situation also maps onto the same ``collapse'' described by Budish et al. The increasing rise of stablecoins perhaps offers an opportunity to analyse the risk trade-offs, if not the confidentiality trade-offs, in greater detail, even if the stablecoins are not exactly cash-like. 

\subsection{Cryptoassets and credit}

The properties of identification and intermediation Kahn et al. describe in their definition of money are core to the development of our argument that the ledger-based record keeping, public visibility and infallible memory of blockchains render cryptoassets more like credit, per their definition.

This strengthens our prior assertion that while it is possible to map cryptoassets onto a slightly modified version of the theft-of-goods model (Section 3.1), a more correct analogy would be drawn to model 3.3, \textit{Money as an Alternative to Full Information}, where the cryptoassets represent a simplified form of credit. An individual who defects can have their credit revoked individually, and system-wide collapse in confidence in credit results in a reversion to using money. 

However, in most current public permissionless ledgers, money does not exist in the form envisaged by Kahn et al. Stablecoins on a public, permissionless ledger are not private enough to qualify; additionally, most exist on a ledger with an endogenous token. If the endogenous token---analagous to credit---collapses, then the stablecoin is unlikely to escape collapse. Though the collapse of tightly paired endogenous tokens and stablecoins, like UST and Luna, has occurred, the fact that UST was not exogenously secured means that for now, we must confine this to being a hypothesis. 

Though cryptoassets arguably have a form (as ``tokens'') and credit per se does not, credit is typically represented on a ledger with (ideally) infallible memory, such that the distinction between the two becomes murky.\footnote{It should be noted that Kocherlakota describes money as ``imperfect memory'' snapshotting a state in the absence of ``costless access to a historical record of past actions'' for agents.~\cite{KOCHERLAKOTA1998232} We, like Kahn et al., argue that it is exactly the absence of memory that gives money its definition, and utility.} Many ledgers are implemented using an account-to-balance mapping, such that the tokens themselves are not atomic and do not exist independently of their system of record (in this case, the ledger). They are both abstract and fully fungible. Thus this makes them incompatible with an analogy to a physical good, but compatible with the implied accounting mechanism for credit in the Kahn model, which is implicitly a global ledger, since all agents can see defections, such as failure to supply. 

In our classification, we argue these public stablecoins' security model is still endogenous, even if their trust locus is exogenous. This means that for the purposes of vulnerability to attack, they are more similar to most endogenous cryptoassets than physical cash. Their fungibility can be limited. The same is not true of cash. A cash-like asset could be in same quadrant as an endogenous token, depending on its representation. However, as described by Kahn et al., confidentiality is inherent to the definition of cash, which in many cases immediately rules out any equivalence between stablecoins on public ledgers and cash. As stated in the title of this paper, privacy is fungibility. 

\section{Conclusion}

The proliferation of stablecoins could be seen as a solution to ledgers needing a cash-like primitive. However, account abstractions, public visibility and other limitations mean that stablecoins typically exist in the endogenous/endogenous quadrant, unless run on a private system with an exogenous trust locus.

By analysing the models of Kahn et al., we can see that cryptoassets map most closely onto the situation described by a simplified credit model rather than by their model for money. Indeed, if we extend their model to create a simplified model for the theft of assets in a situation of full knowledge and full expected utility on a ledger with endogenous tokens, we can intuit that theft should happen extremely frequently. However, the security mechanisms of a ledger, that is, collapse of trust and its endogenous token, will not occur for the reason that trust in the governance topology of a ledger is unlikely to be undermined by a single theft, unless it is particularly large, or affects an important account, such as the ledger's Foundation or Community Pool. However, in those cases, we would expect such an attack to be reverted via governance.

Thus we argue that the network topology of a ledger is sufficient to guard against the most common forms of opportunism by hostile agents. This leaves users at risk mainly to the same kind of low-effort attacks that occur in other online spaces, for example phishing. We do note, however, that large holders are subject to a higher level of physical risk, due to the same property Kahn identified in his paper---that in order to steal credit, you need to steal identity first. This is often easier to do via physical than by technical means. Larger holders are also more likely to be targeted by network governance, suggesting that there is a relationship between holding and risk that is increased as holdings increase, independent of the hypothetical attacker (criminals or network governance).

In this paper, we have argued for the utility of a cash-like asset on permissionless networks in order for users to transact. Nevertheless, even a confidential asset on a public, permissionless network could be vulnerable to a simpler attack---halting the ledger. If the ledger had an endogenous token as well as a cash-like primitive, then to some extent the risk of a halt collapsing the value of an endogenous token and thus destroying the network is shared by any assets that also rely on the ledger. This means that global ledger state implies a single point of failure, and that privacy alone is not enough to make a cash-like asset function optimally. 

Considering this another way, the public, permissionless ledger is an intermediary of sorts if it is both the authoritative record, and has agency. Of course, it is possible to ``switch off'' cash backed by private banks and a government, but there is a system-level cost. The real question perhaps, is whether the exogenous cost of doing so is formally different from that of switching off or halting a ledger.

If indeed blockchains are emergent institutions,~\cite{DAVIDSON_DE_FILIPPI_POTTS_2018} then absent a formal difference in exogenous cost, we might assume that operating with an oblivious asset rather than a public, endogenous asset would result in the systemic properties described by Kahn et al.

\bibliographystyle{splncs04}
\bibliography{biblio}


\appendix



%
%
%

\end{document}